\documentclass[aps,amsfonts,doublespace]{revtex4}
\usepackage{epsfig,amsopn}
\usepackage{graphicx}

\begin{document}

\def\be{\begin{equation}}
\def\ee{\end{equation}}
\def\bea{\begin{eqnarray}}
\def\eea{\end{eqnarray}}
\def\bml{\begin{mathletters}}
\def\eml{\end{mathletters}}
\def\l{\label}
\def\b{\bullet}
\def\eqn#1{(~\ref{eq:#1}~)}
\def\no{\nonumber}
\def\av#1{{\langle  #1 \rangle}}

\title{Exact and limit distributions of the largest fitness on correlated fitness landscapes}

\author{Kavita Jain$^1$, Abhishek Dasgupta$^2$  and Gayatri Das$^3$}
\email{jain@jncasr.ac.in, abhidg@iiserkol.ac.in, gayatridas@jncasr.ac.in}
\affiliation{$^1$Theoretical Sciences Unit and Evolutionary and
  Organismal Biology Unit, Jawaharlal Nehru Centre for Advanced
  Scientific Research,  Jakkur P.O., Bangalore 560064, India}
\affiliation{$^2$Indian Institute of Science Education and Research, Block HC, Salt Lake City, Kolkata 700106, India}
\affiliation{$^3$Theoretical Sciences Unit, Jawaharlal Nehru Centre for Advanced Scientific Research,  Jakkur P.O., Bangalore 560064, India}
\widetext
\date{\today}

\begin{abstract}
We study the distribution of the maximum of a set of random 
fitnesses with fixed number of mutations in a model of biological 
evolution. The fitness variables are not independent and the
correlations can be varied via a parameter $\ell=1,...,L$.   
We present analytical calculations for the following three 
solvable cases: (i) one-step mutants with arbitrary $\ell$ (ii) weakly 
correlated fitnesses with
$\ell=L/2$ (iii) strongly correlated fitnesses with $\ell=2$. In all these
cases, we find that the limit distribution for the maximum fitness is not
of the standard Gumbel form.  
\vskip0.5cm
\end{abstract}
\maketitle

{\it Introduction}: Extreme value theory \cite{David:2003,Clusel:2008} has
found applications in various diverse fields ranging from  
physics of disordered systems such as spin glasses \cite{Derrida:1981}
and driven diffusive systems \cite{Jain:2003} to hydrology
\cite{Katz:2002} and finance \cite{Bouchaud:2003}. 
Here we are interested in its role in a model that describes the biological
evolution of an infinitely large population of asexually replicating
genetic sequences. The (logarithmic) 
population of a  sequence increases linearly with 
time with the slope given by the sequence fitness and the intercept by
the number $D=0,...,L$ of mutations with respect to the reference sequence
\cite{Krug:2003}. It has been shown that out of the $S_D={L \choose D}$
sequences present at constant $D$, the population dynamics involve
only the sequence with the 
{\it largest fitness} at given $D$ \cite{Jain:2005}. 

If the sequence 
fitnesses are uncorrelated random variables chosen from a distribution
decaying faster than a power law, the largest fitness is distributed
according to the well known Gumbel distribution \cite{David:2003}.  
However as several experimental and theoretical studies have indicated that
the realistic fitness landscapes are not completely random 
\cite{Gavrilets:2004}, we are led to study the extreme 
statistics of  {\it correlated} fitnesses.  
In recent studies of extreme statistics of strongly correlated
variables, deviations from the Gumbel 
distribution have been shown numerically (see, for example,
\cite{Bolech:2004}) or by  
analysing the tails of the extremal distribution
\cite{Carpentier:2001,Dean:2001} but very few analytical results for
the {\it full distribution} have been obtained 
\cite{Majumdar:2004,Fyodorov:2008}.  In this Letter, we obtain
analytical results for the full distribution 
for both weak and strong correlations and show that it has a 
non-Gumbel form.   

{\it Block model}: 
We consider 
a block model \cite{Perelson:1995} of protein evolution in which 
a protein sequence  of length $L$ is represented by
a binary string of $0$'s and $1$'s and 
divided into $B$ blocks of equal length $\ell=L/B$. The block fitness 
$f_{j}(d)$ gives the fitness of a block with $d$ ones and the $j$th
permutation of such $s_d={\ell \choose d}$ possible random variables,
each of which are chosen independently from a common exponential 
distribution. The sequence fitness is given by the average of the
corresponding block fitnesses and two sequence fitnesses are
correlated when they share at least one block fitness.  
An attractive feature of the block model is that the correlations amongst the fitnesses and the structure of the fitness landscape can 
be {\it tuned} with the block length $\ell$. For $\ell=1$, the
sequence fitnesses are strongly correlated and the fitness landscape
is smooth, while for 
$\ell=L$, the model has uncorrelated fitnesses and the fitness landscape is maximally rugged. 
In the following, we work with even $L$ 
and consider $D \leq L/2$ as  the results for $D > L/2$ can be
obtained on simply replacing $D$ by  $L-D$. The number $D$ of
mutations is measured with respect to the reference sequence
$\{0,0,...0\}$. 
  
{\it One-step mutants, any $\ell$}: We first consider the extremal
distribution for the fitnesses which carry only one mutation as this
case can be solved for any $\ell$.  
Although there are $L$ one-step mutants, the number of sequences with
distinct fitness is $\ell$ as the fitness $w_j$ of one-mutant neighbor
is given by 
\be
w_j=\frac{(B-1) f_{1}(0)+f_{j}(1)}{B} ~,~j=1,...,\ell \l{d1defn}
\ee
\begin{figure}
\begin{center}
\includegraphics[angle=270,scale=0.4]{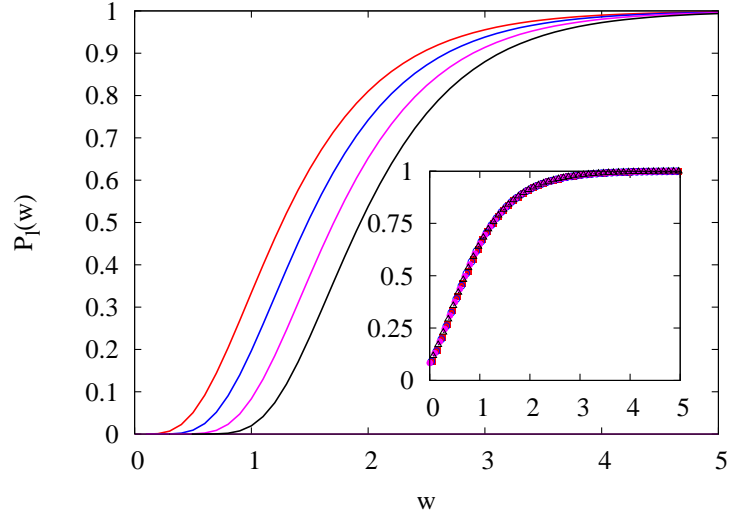}
\caption{Cumulative distribution ${\cal P}_l(w)$ for one-step mutants shown for
  $B=3$ and $\ell=5, 10, 20, 40$ (left to right) calculated using
  (\ref{d1expr1}). The inset shows the data collapse when the
  distribution is plotted as a function of $u=w-B^{-1} \ln l$.}
\label{d1fig}
\end{center}
\end{figure}
Since the cumulative distribution ${\cal P}_\ell(w, D, L)$ gives the
probability that all the $w_j$'s are smaller than $w$, we have 
\bea
{\cal P}_\ell(w,1, L) = \int_0^\infty df_1(0) e^{-f_1(0)} \prod_{j=1}^\ell
\int_0^\infty df_j(1)
e^{-f_j(1)} \Theta(w-w_j)  
= \int_0^{\frac{B w}{B-1}} df  e^{-f} \left[1-e^{-B w}
  e^{(B-1) f} \right]^\ell \l{d1expr1}
\eea
where $\Theta(..)$ is the Heaviside theta function. 
The cumulative distribution calculated using the
above equation is shown in 
Fig.~\ref{d1fig} for various $\ell$. For $\ell=1$, the distribution
${\cal P}_1(w)=1+(L-2)^{-1} \left[e^{-L w}- (L-1) e^{-L w/(L-1)} \right]$ while the
double exponential form $e^{-L
  e^{-w}}$ is obtained for $\ell=L$.  
Thus we have an example of a family of extremal distributions that
interpolates between exponential and Gumbel distributions as
correlations are varied.

The integral on the right hand side (RHS) of (\ref{d1expr1}) does not
seem to be exactly doable but for fixed $B$, it is possible to 
cast it in a scaling form which turns out to be of non-Gumbel 
form.  Following an integration by
parts, (\ref{d1expr1}) can be rewritten as  
\bea
{\cal P}_\ell(w, B) &=&(1-e^{-B w})^{\ell}-\ell e^{-\frac{B w}{B-1}}
\int_0^{1-e^{-B w}} dz ~z^{\ell-1} (1-z)^{\frac{-1}{B-1}} \l{d1expr2}\\
&=& (1-e^{-B w})^{\ell} e^{-\frac{B w}{B-1}} \sum_{n=0}^\infty
 {\frac{2-B}{B-1}+n \choose n} \frac{n}{\ell+n} (1-e^{-B w})^n~,~B > 1
 \l{d1expr3} 
\eea
It is evident from the last expression that in the limit $\ell, w \to
\infty$ with $\ell e^{-B w}$ finite, ${\cal P}_\ell(w, B)$ deviates from Gumbel
distribution for $B > 1$. Since the summand in (\ref{d1expr3}) peaks
around $n \sim  e^{B w} \gg 1$ as $w \to \infty$, the binomial
coefficient can be approximated by
$n^{\frac{2-B}{B-1}}/\Gamma(\frac{1}{B-1})$ for large $n$. Replacing the 
sum in (\ref{d1expr3}) by an integral and defining the scaling variable $u=w- B^{-1} \ln
\ell$ , we finally have  
\be
{\cal P}_\ell(w, B) \approx \frac{e^{-\ell e^{-B
      w}}}{\Gamma(\frac{1}{B-1})} \int_0^{\infty} dn~ 
\frac{n^{\frac{1}{B-1}} e^{-n}}{\ell e^{-B
    w}+n} 
 = \frac{e^{\frac{-B u}{B-1}}}{B-1} \Gamma\left(\frac{1}{1-B},e^{-B u}
 \right)
\l{gamma}
\ee
where the last expression holds for all $u$ except $u \to \infty$ and  $\Gamma(a,x)$ is the incomplete gamma function
\cite{Gradshteyn:1980}. Thus the 
limit distribution ${\cal P}_\ell(w, B)$  is a function of $u$ (see inset) and is  of {\it traveling wave} form $F_B(w-v t)$ if we
identify $t$ by $\ln \ell$ and velocity $v$ by $B^{-1}$ (also see
(\ref{Bessel}) below). Note that unlike previous works
\cite{Dean:2001,Fyodorov:2008} that assume the
distribution to be of traveling wave form, here we have shown the existence
of such a solution. 


{\it Block length $\ell=L/2$, any $D$}: As $B$ is an
integer, $L/2$ is the largest value of $\ell$ at which correlations
are nonzero. We now turn 
to this case with weak correlations and show that the
distribution is of non-Gumbel form for any $D$. For $B=2$, the fitness of a
sequence with $D$ ones can be obtained by averaging over the block fitnesses with $d'$ ones in the 
first block and $d''=D-d'$ ones in the second block. As there are $s_{d'}$ possible fitnesses for the first block and $s_{d''}$ for the second, the sequence fitness takes the following form:
\be
w_{j,k}(d')=
\frac{f_j(d')+f_k(d'')}{2}~,~d'=0,...,d_{u}~,~j=1,...,s_{d'}~,~k=1,...,s_{d''}\l{lhalfdefn}
\ee 
where $d_u=(D-1)/2$ for odd $D$ and $D/2$ for even $D$. The above equation gives distinct $w_{j,k}(d')$ for all $d'$ except $d'=D/2$ for which as  $d'=d''$, 
distinct fitnesses are obtained when the index $k$ runs from $j$ to $s_{D/2}$. Thus
the number of distinct random variables are given
by $(1/2) \left({2 \ell \choose  D}+
{\ell \choose D/2} (1-D ~\textrm{mod} ~2) \right)$
\cite{Gradshteyn:1980} which increases as $\sim \ell^D$. As we shall see below, the extreme value distribution depends on whether $D$ is odd or even. 

\noindent(i) For odd $D$, the fitnesses $w_{j,k}(d')$ are identically
distributed as is evident from (\ref{lhalfdefn}). The probability that
all the fitnesses are smaller than $w$ is given by 
\be
{\cal P}_{L/2}(w,L) =  \prod_{d'=0}^{d_{u}} \prod_{j=1}^{s_{d'}}
\prod_{k=1}^{s_{d''}} 
\int_0^\infty df_j(d') e^{-f_j(d')} \int_0^\infty df_k(d'') e^{-f_k(d'')}
\Theta(w-w_{j,k}(d'))
\l{lhalfodd1}
\ee
In the above expression, the product $\prod_{j=1}^{s_{d'}} \Theta(w-w_{j,k}(d'))$ in the integral over $f_k(d'')$ 
 requires that $f_k(d'') < 2 w- f_j(d')$ for all $j=1,...,s_{d'}$.  
 It is however sufficient to satisfy $f_k(d'') < 2 w- f_J(d')$ where $f_J(d')=\max
\{f_1(d'),...,f_{s_{d'}}(d')\}$.  Furthermore, as $f_k(d'')$ is positive, $2 w-f_{j}(d')$ must also be positive for all $j$ thus restricting the domain of integration over $f_j(d')$ to $2 w$. Thus we can write 
\bea
\int_0^{\infty} df_k(d'') e^{-f_k(d'')} \prod_{j=1}^{s_{d'}} \Theta(2
w-f_{j}(d')-f_k(d'')) 
= \sum_{J=1}^{s_{d'}} \Theta(2 w- f_J(d')) \prod_{\stackrel{j=1}{j
    \neq J}}^{s_{d'}}
\Theta(f_J(d')-f_{j}(d')) (1-e^{-2 w+ f_J(d')})
\eea
which is independent of $k$. As a result, the product over $k$ in
(\ref{lhalfodd1}) can be done using the basic properties of Heaviside theta function. This immediately gives 
\bea
{\cal P}_{L/2}(w,L) 
&=& \prod_{d'=0}^{d_u} \sum_{J=1}^{s_{d'}} \int_0^{2 w} df_J(d')
e^{-f_J(d')} (1-e^{-2 w+
  f_J(d')})^{s_{d''}} \prod_{\stackrel{j=1}{j
    \neq J}}^{s_{d'}} \int_0^{f_{J}(d')}
 df_j(d') e^{-f_j(d')} \\
&=& \prod_{d'=0}^{d_u} s_{d'} \int_0^{2 w} df ~e^{-f}
(1-e^{-2w+f})^{s_{d''}} (1-e^{-f})^{s_{d'}-1} \l{lhalfodd22}
\eea
For $D=1$, the above expression reduces to (\ref{d1expr1}) with
$\ell=L/2$. Following the steps similar to those leading to
(\ref{d1expr3}), we rewrite the last equation as 
\bea
{\cal P}_{L/2}(w,L) 
&=& \prod_{d'=0}^{d_u} s_{d'} s_{d''} (1-e^{-2 w})^{s_{d'}+s_{d''}}
\sum_{n=0}^{\infty} \frac{(1-e^{-2 w})^n}{(n+s_{d'}) (n+s_{d''})} ~\frac{(n+s_{d''})!
  (n+s_{d'})!}{n! (n+s_{d'}+s_{d''})!}
\l{lhalfodd2}
\eea
To find the limit distribution, we first note that the
factor corresponding to $d'=0$ in (\ref{lhalfodd22}) is of the 
form (\ref{d1expr1}). On comparing, we infer the scaling variable
for $d'=0$ term to be $s_D e^{-2 w} \sim \ell^D e^{-2 
    w}$ when $\ell, w \to  \infty$.  This suggests that for
arbitrary $d'$, the product $s_{d'} s_{d''} e^{-2 w}$ remains finite
while $s_{d'} e^{-2 w}, s_{d''} e^{-2 w} \to 0$ for large $\ell$ and
$w$. In these scaling limits, for large $n$, we can write 
\be
\frac{(n+s_{d''})!
  (n+s_{d'})!}{n! (n+s_{d'}+s_{d''})!} \approx \left(
  \frac{n+s_{d'}}{n+s_{d'}+s_{d''}} \right)^{s_{d'}} \approx e^{-s_{d'}
  s_{d''}/n}=\textrm{exp}\left[-{e^{-2 u_{d'}}}/n e^{-2 w}\right]
\ee
where $u_{d'}=w- \ln (\sqrt{s_{d'} s_{d''}})$. Approximating the sum 
in (\ref{lhalfodd2}) by an integral, we finally get 
\bea
{\cal P}_{L/2}(w,L) \approx   \prod_{d'=0}^{d_u} e^{-2 u_{d'}}
\int_0^{\infty} \frac{dn}{n^2}~e^{-n} ~e^{-\frac{e^{- 2 u_{d'}}}{n}}= 
  \prod_{d'=0}^{d_u} 2 e^{-u_{d'}} K_1(2 e^{-u_{d'}}) 
\l{Bessel}
\eea
where $K_n(x)$ is the modified Bessel function of the second
kind \cite{Gradshteyn:1980}.  Interestingly, the above distribution for
$D=1$ has the same form as the cumulative
distribution for the minimum energy in a random energy model with
logarithmically correlated potential \cite{Fyodorov:2008}. However, for $D >
1$, there does not appear to be a single scaling variable. 

\noindent(ii) For even $D$, since the fitnesses
$w_{j,j}(D/2), j=1,...,s_{D/2}$ have a different distribution than
the rest,  the fitnesses are not identically distributed in this case. 
Using the results obtained above for $d' < D/2$ and
separating the contribution due to $d'=D/2$, we can write 
\be
{\cal P}_{L/2}(w,L)=\prod_{d'=0}^{d_u-1} s_{d'} \int_0^{2 w} df ~e^{-f}
(1-e^{-2w+f})^{s_{d''}} (1-e^{-f})^{s_{d'}-1} \times
\prod_{j=1}^{s_{D/2}} \prod_{k=j}^{s_{D/2}} \int_0^{\infty}
 df_j(D/2) e^{-f_j(D/2)}  \Theta(w-w_{j,k}(D/2))
\ee
By applying the same procedure as for odd $D$, the integral over
$f_j(D/2)$ can be evaluated to give $1-e^{-w}$. Within the same
scaling limits as for odd $D$, we  finally obtain
\be
{\cal P}_{L/2}(w,L)
\approx e^{-e^{-u_{D/2}}}\prod_{d'=0}^{d_u-1} 2 e^{-u_{d'}} K_1(2 e^{-u_{d'}}) 
\ee

{\it Block length $\ell=2$, any $D$}: For $\ell=2$, although the
sequence fitnesses are not only strongly correlated but
non-identically distributed as well, it is possible
to solve for the extreme value distribution exactly.  If $n_1$ and $n_2$ denote the number of blocks with fitness $f_1 (1)$
and $f_2 (1)$ respectively, the number of blocks $n_3$ with fitness $f_1(2)$ at a fixed $D$ is given by $(D-n_1-n_2)/2$. Furthermore, as the total number of blocks equals $B$, there are $(L-D-n_1-n_2)/2$ number of blocks with fitness $f_1(0)$. Thus the fitness $w_{n_1,n_2}$ of a sequence
with $D$ mutations obtained by averaging over the block fitnesses is writeable as 
\be
w_{n_1,n_2}= \frac{(L-D-n_1-n_2) f_1(0)+ 2 n_1 f_1 (1)+ 2 n_2
  f_2 (1)+ (D-n_1-n_2) f_1(2)}{L}
\ee
The cumulative distribution ${\cal P}_2(w, L)$ is given by
\be
{\cal P}_2(w, L)= \int_0^{\infty} ... \int_0^{\infty} df_1(0) df_1(1)
df_2(1) df_1(2) 
e^{-f_1(0)-f_1(1)-f_2(1)-f_1(2)} \prod_{n_1,
  n_2=n_{1,l},n_{2,l}}^{n_{1,u},n_{2,u}} 
\Theta(w-w_{n_1,n_2})  
\l{l2general}
\ee
where $n_{i,u} (n_{i,l})$ is the maximum(minimum) allowed value of
$n_i, i=1, 2$ which, as discussed below, depends on whether $D$ is odd or even. Before proceeding further, we first note that in the product over theta functions
in the above integrand, only those factors in
which at least one of the indices $n_1, n_2$ are zero need to be
retained and the rest are redundant.  To see this, consider the theta functions with a given $n_1+n_2$. Then if $f_1(1) >
f_2(1)$, the fitness 
$w_{n_1,n_2} < w_{n_1+n_2,0} $ so that the condition 
$\Theta(w-w_{n_1,n_2})$ is automatically
satisfied by $\Theta(w-w_{n_1+n_2,0})$. Similarly if $f_2(1) 
>  f_1(1)$, it is enough to keep $\Theta(w-w_{0,n_1+n_2})$.

\noindent(i) For even $D$,  as $n_3$ is an integer, both $n_1,
n_2$ should be either odd or even which implies $n_{i,u}=D,
n_{i,l}=0$. Besides, the conditions $n_1 +n_2 \leq D, n_1 \leq 
D$ should be satisfied as $n_3$ is nonnegative. Counting the number of
possibilities, we find that the total  
number of distinct fitnesses increases as $((D+2)/2)^2$ for $D \leq
L/2$. Using the redundancy argument given above in
(\ref{l2general}), we have 
\be
{\cal P}_2(w, L)= \int_0^{\infty} \int_0^{\infty} df_1(0)  df_1(2) e^{-f_1(0)-f_1(2)}
\Theta(w-w_{0,0}) \left[\int_0^{\infty} df_1(1) e^{-f_1(1)} \prod_{n_1=1}^{D}
  \Theta(w-w_{n_1,0}) \right]^2
\ee
It is easy to see that the integral
over $f_1(1)$ is nonzero provided $f_1(1) < \min \{\alpha+ \beta,...,
(\alpha/D)+ \beta \}$ where we have defined  $\alpha=(L w-(L-D)
f_1(0)-D f_1(2))/2$ and $\beta=(f_1(0)+f_1(2))/2$.
For $\alpha > 0$, this condition reduces to $f_1(1) < (\alpha/D)+\beta$ 
while for $\alpha < 0$, we require $f_1(1) < \alpha+\beta$. Thus we obtain
\be
\int_0^{\infty} df_1(1) e^{-f_1(1)} \prod_{n_1=1}^{D}
  \Theta(w-w_{n_1,0})= \Theta(\alpha)
  \Theta \left(\frac{\alpha}{D}+\beta \right) (1-
  e^{-\frac{\alpha}{D}-\beta})+\Theta(-\alpha) \Theta
  \left(\alpha+\beta \right) (1- e^{-\alpha-\beta}) 
\ee
Using $\Theta(w-w_{0,0})=\Theta(\alpha)$, we finally get
\bea
{\cal P}_2(w, L) = \int_0^{\infty} \int_0^{\infty} df_1(0)  df_1(2) e^{-f_1(0)-f_1(2)}
\Theta(\alpha)  (1-e^{-\frac{\alpha}{D}-\beta})^2 
= \int_0^{\frac{w}{1-y}} df e^{-f} (1-
e^{-\frac{w- (1-2 y) f}{2 y}})^2 (1-e^{-\frac{w- (1-y)
    f}{y}}) 
\l{l2deven1}
 \eea
where $y=D/L$. The above integral can be easily computed and an
explicit {\it exact expression} for the distribution
$P_2(w,L)=d{\cal P}_2(w, L)/dw$ is given by 
\be
P_2(w, L)=\frac{-2 e^{-\frac{w}{2 y}}}{1-4 y}+
\frac{2 e^{-\frac{3 w}{2 y}}-e^{-\frac{2
      w}{y}}+e^{-\frac{w}{(1-y)}}}{1-2 y}+\frac{4 y e^{-\frac{3 w}{2 (1-y)}}}{1-6
  y+8 y^2}+\frac{y (e^{-\frac{w}{y}}-e^{-\frac{2 w}{(1-y)}})}{1-5 y+6 y^2}
\ee
The mean $\bar{w}$ and the variance $\sigma^2$ calculated using
$P_2(w)$ are then given by 
\bea
\bar{w} = 1 + \frac{55}{36} y ~,~\sigma^2 = 1- \frac{3804}{1296} y+
\frac{8135}{1296} y^2 
\eea
Thus the mean increases linearly with $y$ but the variance varies
non-monotonically - it initially decreases with $y$ 
and then increases with the minimum at $y^*=3804/16270 \approx 0.233$.

\noindent(ii) If $D$ is odd, we require that either $n_1$ is odd and 
$n_2$ is even or viceversa alongwith the condition $n_1+n_2 \leq D, n_1
\leq D$. In this case, $n_{i,u}=D, n_{i,l}=1$ and we obtain $(D+1)
(D+3)/4$ distinct fitnesses 
for $D \leq L/2$. Following the same reasoning as above, the
cumulative distribution for odd $D$ can be written as 
\bea
{\cal P}_2(w, L) &=& \int_0^{\infty} \int_0^{\infty} df_1(0)  df_1(2) e^{-f_1(0)-f_1(2)}
\left[\int_0^{\infty} df_1(1) e^{-f_1(1)} \prod_{n_1=1}^D
  \Theta(w-w_{n_1,0}) \right]^2 \\
&=&\int_0^{\infty} \int_0^{\infty} df_1(0)  df_1(2) e^{-f_1(0)-f_1(2)}
\left[ \Theta(\alpha)\Theta
  \left(\frac{\alpha}{D}+\beta \right) (1-
  e^{-\frac{\alpha}{D}-\beta})^2 
  +\Theta(-\alpha) \Theta
  \left({\alpha}+\beta \right) (1-
  e^{-{\alpha}-\beta})^2 \right]  \l{l2dodd1}
\eea
The first term in the above sum  reduces to (\ref{l2deven1}). 
In the second term, the condition $\Theta \left(\alpha+\beta \right)$ 
requires that $0< (D-1) f_1(2) < L w-(L-D-1) f_1(0)$ and the condition 
$\Theta(-\alpha)$ can be satisfied if (i) $D f_1(2) > L w- (L-D) f_1(0)
> 0$ and (ii) $D f_1(2) > 0 ~,~ L w- (L-D) f_1(0) < 0$. 
Putting all these conditions together, the second term can be evaluated. However for $L \gg 1$, the contribution of the second term to (\ref{l2dodd1}) 
can be neglected and we obtain the same result as for even $D$. 

{\it Conclusions:} We have presented several analytical results for the
extreme distribution in a model with tunable correlations. When the
fitnesses are strongly correlated and non-identically 
distributed, full distribution is obtained exactly. 
For $D=1$ and arbitrary $\ell$, we have shown that the limit
distribution is of traveling wave form. As the limit distribution in
the large $\ell$ limit for $D=1$ and $\ell=L/2$ for any $D$ are seen to
be of traveling wave form, we expect that this form survives for
$\ell \sim {\cal O}(L)$ and any $D$. The weakly
correlated  
model with $D=1$ is found to obey the same extreme statistics as a
random energy model with correlations. An elucidation of the
connection between these two apparently unrelated models would be
interesting.

{\it Acknowledgements:} K.J. thanks S. Sabhapandit for useful comments on the manuscript. 
A.D. thanks JNCASR for hospitality during his stay under the Summer
Research Fellowship Program 2009. 


\end{document}